\documentclass{aa}
\usepackage{times}
\usepackage{graphicx}
\def\kms{\ifmmode {\,{\rm km\,s^{-1}}}                          
        \else {\hbox{$\,$ {\rm km$\,$s$^{\rm -1}$}}}\fi}
\def\solar {\ifmmode_{\mathord\odot} \else $_{\mathord\odot}$\fi} 
\def\msun {\ifmmode {\,{\it M}\solar} \else $\,M$\solar\fi}     
\def\lsun {\ifmmode {\,{\it L}\solar} \else $\,L$\solar\fi}     
\def\half{\ifmmode \textstyle{1\over2} \else $\textstyle{1\over2}$\fi} 
\def\as {\ifmmode {^{\scriptscriptstyle\prime\prime}}           
        \else $^{\scriptscriptstyle\prime\prime}$\fi}
\def\am {\ifmmode {^{\scriptscriptstyle\prime}}                 
        \else $^{\scriptscriptstyle\prime}$\fi}
\def\deg {\ifmmode^\circ\else$^\circ$\fi}                       
\def\raw {\ifmmode\rightarrow\else$\rightarrow$\fi}             
\def\x  {\ifmmode\times\else$\times$\fi}                        
\def\gsim {\ifmmode {\buildrel>\over\sim}               
        \else {\lower.6ex\hbox{$\buildrel>\over\sim$}}\fi}
\def\lsim {\ifmmode {\buildrel<\over\sim}               
        \else {\lower.6ex\hbox{$\buildrel<\over\sim$}}\fi}
\def\sup#1{\ifmmode {^{\rm #1}} \else $^{\rm #1}$\fi}   
\def\ra[#1 #2 #3.#4]{#1\sup{h}#2\sup{m}#3\sup{s}\llap.#4}       
\def\dec[#1 #2 #3.#4]{#1\deg#2\am#3\as\llap.#4}                 
\def\rax[#1 #2 #3]{#1\sup{h}#2\sup{m}#3\sup{s}}         
\def\decx[#1 #2 #3]{#1\deg#2\am#3\as}                   
\graphicspath{{FIGURES/}}

\begin{document}

\title{Dust emission from young outflows: the case of L\,1157}

\author{F.~Gueth\inst{1}, R.~Bachiller\inst{2}, and M.~Tafalla\inst{2}}

\authorrunning{F.~Gueth et al.}

\institute{$^1$Institut de Radioastronomie Millim\'etrique (IRAM),
300 rue de la Piscine, 38406 Saint Martin d'H\`eres, France\\
$^2$IGN Observatorio Astron\'{o}mico Nacional, Apartado 1143,
28800 Alcal\'{a} de Henares, Spain}

\offprints{F.~Gueth\\ e-mail: gueth@iram.fr}

\date{Received date / Accepted date}

\abstract{We present new high-sensitivity 1.3~mm bolometer
observations of the young outflow L\,1157. These data show that the
continuum emission arises from {\em four} distinct components: a
circumstellar disk, a protostellar envelope, an extended flattened
envelope --the dense remnant of the molecular cloud in which the
protostar was formed--, and the outflow itself, which represents
$\sim$20\% of the total flux. The outflow emission exhibits two peaks
that are coincident with the two strong shocks in the southern lobe of
L\,1157. We show that the mm continuum is dominated by thermal dust
emission arising in the high velocity material. The spectral index
derived from the new 1.3~mm data and 850~$\mu$m observations from
Shirley et al.\ (2000), is $\sim$5 in the outflow, significantly
higher than in the protostellar envelope ($\sim$3.5). This can be
explained by an important line contamination of the 850~$\mu$m map,
and/or by different dust characteristics in the two regions, possibly
smaller grains in the post-shocks regions of the outflow. Our
observations show that bipolar outflows can present compact emission
peaks which must not be misinterpreted as protostellar condensations
when mapping star forming regions.}


\maketitle

\section{Introduction} 

Mm and submm mapping of the dust thermal continuum emission has proven
to be a powerful tool to study the density structure of low-mass
protostellar condensations at different evolutionary stages (e.g.\
Motte et al.\ 1998; Chandler \& Richer 2000; Hogerheijde \& Sandell
2000; Shirley et al.\ 2000; Motte \& Andr\'e 2000). However, virtually
all protostars are driving powerful molecular outflows (Bachiller
1996), which are much more extended than the protostellar envelope
size (e.g.\ Andr\'e et al.\ 1999). Strong interaction between the
outflow and the envelope is observed in a few cases (e.g.\ L\,1527,
Motte \& Andr\'e 2000). But besides such local perturbations, there is
increasing observational evidence that the outflow itself can be
responsible for (sub)mm continuum emission also at long distances from
the driving source (Shirley et al.\ 2000; Chini et al.\ 2001). Since
such continuum emission can be a valuable tracer of the outflow
material, its detailed origin and structure clearly deserves further
study.
 
The outflow located in the L\,1157 molecular cloud, at a distance of
440~pc, is one of the few which have been detected so far in the
continuum submm and mm emission (Shirley et al.\ 2000; Chini et al.\
2001). This outflow is driven by an extremely young Class~0 source
(Gueth et al.\ 1997). The southern lobe consists of two large
overlapping cavities, each one ended by a strong bow shock, which are
misaligned due to the probable precession of the ejection axis (Gueth
et al.\ 1996, 1998; Bachiller et al. 2001). The two shocks exhibit a
rich molecular composition (Tafalla \& Bachiller 1995; Zhang et al.\
1995; Avery \& Chiao 1996; Bachiller \& P\'erez-Guti\'errez 1997)
which makes L\,1157 the prototype of chemically active outflows
(Bachiller et al.\ 2001). In this letter, we report new
high-sensitivity bolometer observations of L\,1157 at mm wavelengths,
which, in combination with previously published maps, provides basic
information on the continuum emitting regions.

\begin{figure*}[t]
\includegraphics[angle=270,width=14.5cm]{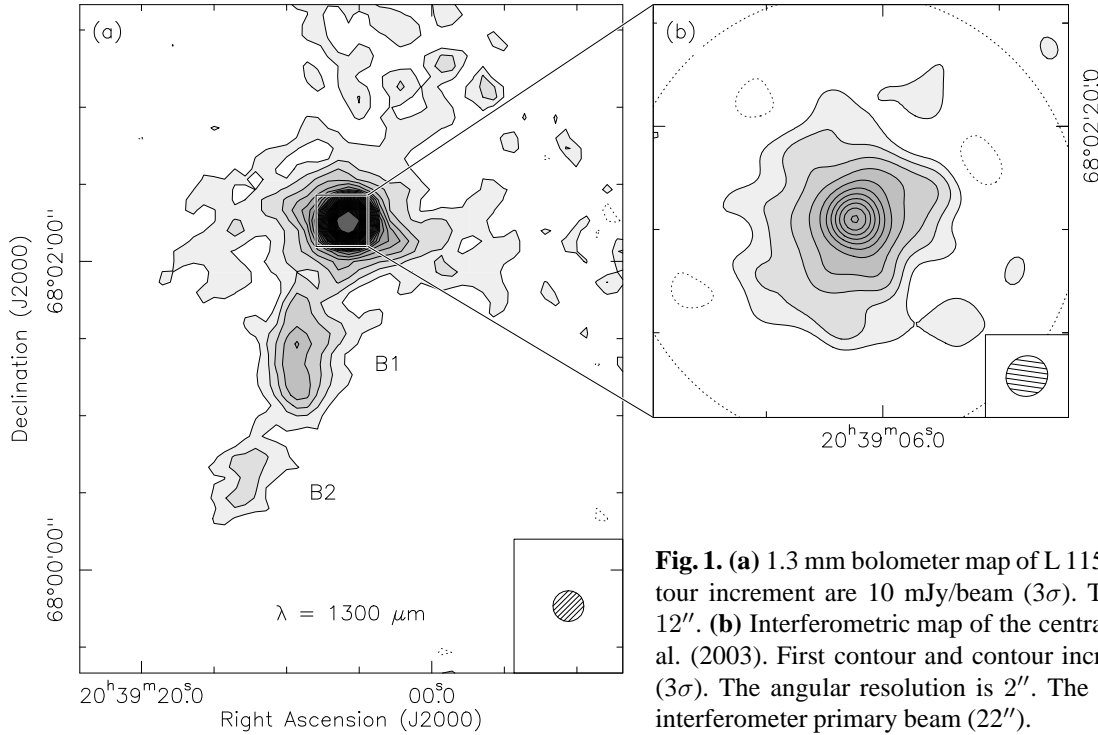}\hspace{-5.9cm}\parbox[t]{9.4cm}{
\vspace{7cm}\caption{{\bf (a)} 1.3~mm bolometer map of L\,1157. First
contour and contour increment are 10~mJy/beam (3$\sigma$). The angular
resolution is $12''$. {\bf (b)} Interferometric map of the central
region, from Beltr\'an et al.\ (2003). First contour and contour
increment are 10~mJy/beam (3$\sigma$). The angular resolution is
$2''$. The dash circle indicates the interferometer primary beam
(22$''$).}}
\label{fig.bolo}
\end{figure*}

\section{Observations} 


The observations were carried out in 1997 February and 1999 December,
with the IRAM 30--m telescope. The detector was the MPIfR 1.3~mm
bolometer array (MAMBO), which had 19 channels in 1997 and 37 channels
in 1999. Each channel has an equivalent bandwidth of $\sim$70~GHz. 8
individual on-the-fly maps were obtained, with a different scanning
direction for each coverage, from nearly perpendicular to almost
parallel to the flow axis.  The antenna beam width was $12''$. Skydips
were observed every one to two hours to monitor the atmospheric zenith
optical depth, which was found to be 0.1 to 0.4. Flux calibration was
achieved by observing Uranus. The final flux accuracy is estimated to
be $\sim$30\%. The data were reduced with the IRAM package (NIC),
using the EKH restoration algorithm. Fig.~1a presents the resulting
image. The rms at the map center is $\sim$3.5~mJy/beam.
 

Beltr\'an et al.\ (2003) obtained IRAM Plateau de Bure interferometric
observations of the 1.3~mm continuum emission of the L\,1157
protostar. The bolometer map presented in this paper was used to
derive the short-spacing information and thereby complement the
interferometric dataset. The resulting map, which covers only the
central region of the bolometer image, is shown in Fig.\,1b.

\section{Results} 

\subsection{Four continuum components} 

The data presented in Fig.\,1 show that the mm continuum emission in
L\,1157 arises from {\em four} different components. Their properties
are summarized in Table~1.

{\bf Central source.} The interferometric observations (Fig.\,1b)
reveal the existence of a compact ($<1''$) source, of $\sim$80~mJy,
located at \ra[20 39 06.24], \dec[68 02 15.6] (J2000). Obviously, this
structure is closely associated to the protostar itself, and it is
tempting to identify it with the thermal dust emission of a
circumstellar disk. We refer to Beltr\'an et al.\ (2003) for a more
detailed discussion.

{\bf Envelope.} The central source is surrounded by an extended
emission, that originates from the infalling (Gueth et al.\ 1997;
Mardones et al.\ 1997) protostellar envelope. On the bolometer map,
the peak flux density is 500~mJy/beam, and the flux integrated in a
circle of 40$''$ diameter is 960~mJy. Assuming a typical dust
temperature of 30~K, the corresponding mass would be
$\sim$2.1\,{\msun} (Table~1). At the resolution of these observations
($12''$), the envelope is smooth and symmetric (the contour at half
intensity is nearly round and has a diameter of $\sim$13$''$, cf.\
Fig.~1a).  However, the interferometric map (Fig.~1b) shows that the
perturbation of the envelope by the outflow affects the internal
structure of the envelope, which is no longer isotropic at that
resolution, but extended along the flow axis.

\begin{table}
\begin{tabular}{lccccc}
Component          & Size              & Flux  & Temp.  & Mass    \\
                   &                   & (mJy) & (K)    & (\msun) \\
\hline 
Central source     & $<1''$            & 80    & 30     & 0.15    \\
Envelope           & $40''$            & 960   & 30     & 2.1    \\
Extended envelope  & $140''\times60''$ & 320   & 13     & 2.1    \\
Outflow            & $40''\times100''$ & 335   & 25--60 & 0.3--0.9   \\ 
\end{tabular}
\caption{Properties of the four components of the 1.3~mm continuum
emission in L\,1157. The observed integrated fluxes are converted into
masses assuming optically thin thermal dust emission, a uniform
temperature, and a mass opacity $\kappa = 0.01$\,cm$^{2}$g$^{-1}$.
The outflow emission has not been corrected for the molecular line
contamination (see text).}
\end{table}

\begin{figure*}[t]
\includegraphics[angle=270,width=18cm]{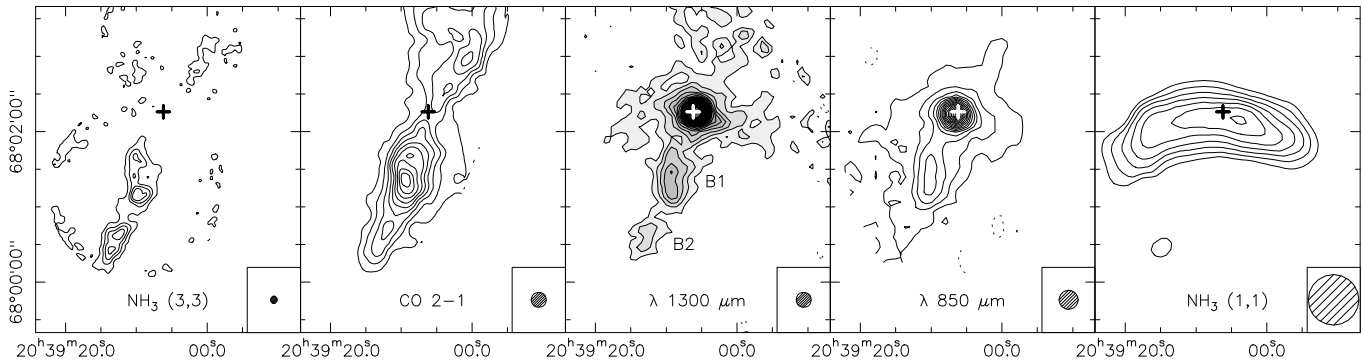}
\caption{Comparison of the 1.3\,mm continuum emission map of L\,1157
with maps of other gas and dust tracers.
{\bf (a)} NH$_3$~(3,3) emission, from Tafalla \& Bachiller (1995).
{\bf (b)} CO~(2--1) emission, from Bachiller et al.\ (2001).
{\bf (c)} $\lambda$1.3~mm continuum emission, see Fig.~1.
{\bf (d)} $\lambda$850~$\mu$m continuum emission, from Shirley et al.\ (2000).
{\bf (e)} NH$_3$ (1,1) emission, from Bachiller et al.\ (1993).}
\end{figure*}

{\bf Extended envelope.} The L\,1157 protostellar envelope is
surrounded by an even more diffuse emission extended roughly
east-west, i.e.\ approximately perpendicular to the flow axis
(Fig.\,1a). As shown in Fig.\,2, this feature is also seen in the
850~$\mu$m continuum map obtained by Shirley et al.\ (2000), as well
as in the lower resolution NH$_3$~(1,1) map obtained by Bachiller et
al.\ (1993). It is also detected in quiescent emission of $^{13}$CO
(Gueth et al.\ 1997), DCO$^+$, and N$_2$H$^+$ (Bachiller et al.\
2001). Very likely this structure is a dense remnant of the molecular
cloud in which the protostar was formed. The size of this envelope in
the mm map is about 140$''\times$60$''$, its flux density is roughly
constant, $\sim$12\,mJy/beam, and its total flux is 320$\pm$50\,mJy
(after removing the 960~mJy corresponding to the central source and
its envelope). If the dust temperature is 13\,K (as suggested by the
NH$_3$ observations of Bachiller et al.\ 1993), the mass of this
extended structure is $\sim$2.1\,{\msun}.

{\bf Outflow.} Besides the emission associated with the protostar
vicinity, Fig.\,1a shows that there is also an emission extending
toward the south-east. A comparison with the \mbox{CO~(2--1)} map of
Bachiller et al.\ (2001), which has the same angular resolution,
clearly shows that this emission is associated to the outflow (see
Fig.~2): the continuum emission presents two main structures (labeled
B1 and B2), whose positions and shapes correspond very well to those
of the brightest CO emission. This is further demonstrated by the
morphological agreement that exists between the mm continuum map and
the NH$_3$~(3,3) emission (Fig.~2) and the high-resolution SiO~(2--1)
maps of Gueth et al.\ (1998). The NH$_3$ and SiO emission trace shocks
in molecular outflows. B1 does also coincide with the HH\,375
Herbig-Haro object (Chini et al.\ 2001). We thus conclude that
significant emission arises from the shocked regions of the L\,1157
outflow. The peak flux densities of B1 and B2 are $\sim$50 and
$\sim$25~mJy/beam, respectively, and their integrated fluxes are 245
and 90~mJy.

\subsection{Origin of the outflow emission}

The broad bolometer bandwidth includes many molecular lines, which are
quite strong in L\,1157, and could thus create a spurious continuum
emission. At the B1 position, the CO~(2--1) line-integrated emission
is 155~K\,km$^{-1}$ (Bachiller et al.\ 2001), which corresponds to
$\sim$8~mJy/beam in the bolometer band, i.e.\ 16\% of the detected
emission. Other lines are weaker, but their sum can however be
important: adding the integrated intensities of all lines (but CO)
detected by Bachiller \& P\'erez-Guti\'errez (1997) at B1 yields
110\,K\,km$^{-1}$. 
This puts the lines contribution to about 30\% of the emission
detected by the bolometer. This value could possibly be somewhat
larger, since the line survey was not complete and not covering the
full bolometer band.

Fig.\,2 also shows the SCUBA 850~$\mu$m map obtained by Shirley et
al. (2000). This map looks very similar to the 1.3\,mm observations.
From Hirano \& Taniguchi (2001), we can roughly estimate the CO~(3--2)
line-integrated emission to be 200\,K\,km$^{-1}$ at the B1 position,
which corresponds to $\sim$20\% of the bolometer measurement. Other
lines must contribute and, as in the 1.3~mm case, could match the CO
intensity and thus raise the line contribution to about 40\%.

We conclude that the lines contributions to the 1.3~mm and 850~$\mu$m
bolometer maps are significant, but are unlikely to account for the
whole emission, which is thus dominated by continuum emission. A
comparison of both maps reveals a spectral index in the range 3 to 5.5
(see below), which is typical for {\em dust thermal emission}, and
thus suggests that any free-free emission is negligible in these
images. Dust is actually thought to play a key role in the chemical
processes taking place in shocked outflows (e.g.\ Schilke et al.\
1997).

It is difficult to estimate the temperature of the dust in the
outflow: the IRAS measurements (that would be needed to build the SED
and then derive a temperature) do not have the angular resolution
required to resolve the different components contributing to the
L\,1157 emission. Tafalla \& Bachiller (1995) used multiline NH$_3$
analysis to derive outflowing gas temperatures of 60 to 80~K. These
values are upper limits to the temperature of the dust, which has a
very short post-shock cooling time (e.g.\ Clark et al.\ 1986). Chini
et al.\ (2001) suggested a dust temperature of 27~K for HH\,375
(B1). From the measured mm fluxes, assuming optically thin emission
and a temperature between 25 and 60~K, we estimate that the mass of
the blueshifted shocked regions is $\sim$0.3 to 0.9\,{\msun}.

\subsection{Spectral index}

Fig.\,3 presents a map of the spectral index $\alpha$ computed between
850~$\mu$m and 1.3~mm: $\alpha$\,=\,log($S_{850\mu{\rm m}}/S_{1.3{\rm
mm}}$)$/$log($1300/850$). This image shows systematic variations
across the mapped region. The lowest values of the index, in the range
3 to 4, are obtained around the protostellar condensation. They are in
good agreement with the average value $\alpha$ = 3.4$\pm$0.3 found by
Shirley et al.\ (2000) in a sample of Class 0 sources. The highest
values are in the range from 4.5 to 5.5, and are measured towards the
outflow region. 

{\bf Possible artifacts.} A calibration error in one or both maps
would translate into an incorrect scale for the spectral indexes, but
would not introduce a position-dependent effect. However, line
contamination of the bolometer measurements in the outflow region can
affect the apparent spectral index by introducing an offset to the
actual index of the continuum emission. Simple algebra shows that the
values discussed in the previous section result in $\alpha$ being
increased by 0.35 at B1, the position where the line contamination
reaches its highest value. The variations of $\alpha$ at other
positions are expected to be smaller. However, uncertainties on the
line contamination, as well as possible calibration errors, make that
this estimate of the $\alpha$ variation is only a rough
approximation. In fact, the offset could be smaller (or even
negative), if e.g.\ line contamination at 1.3~mm was
underestimated. Inversely, if the line contribution at 850~$\mu$m is
higher, the offset can be larger: in order to introduce the observed
1.5 difference between the outflow and the protostellar envelope, a
line/continuum ratio at 850~$\mu$m of $\sim$2 is required. This would
imply that the 850~$\mu$m map is essentially line emission, a
possibility which --from the fluxes of the lines observed in the
band-- seems unlikely.

{\bf Dust properties.} Besides the possible effects of the line
contamination, a significant fraction of the variation in the spectral
index has to be traced back to intrinsic properties of the continuum
emission. For optically thin thermal dust emission, in the
Rayleigh-Jeans limit, $\alpha=2+\beta$, where $\beta$ is the exponent
in the dust opacity law $\kappa\,\propto\,\nu^{\beta}$. Hence, our
observations would suggest that $\beta$ varies substantially from
$\sim$1.5 around the protostar to $\sim$3 in the shocked
regions\footnote{
Note however that the factor $h\nu$/$k$ is $\sim$11\,K at 1.3\,mm and
$\sim$17\,K at 850~$\mu$m, so the Rayleigh-Jeans approximation 
(T$\gg$$h\nu$/$k$) fails e.g.\ in the extended envelope, where T$_{\rm
dust}$$\sim$13\,K (if the dust is at the temperature indicated by the
NH$_3$ observations). In that case, $\alpha = 2+\beta+\gamma(T_{\rm
dust})$, where $\gamma$ can easily be derived from the Planck function
(see e.g.\ Visser et al.\ 1998). At 13~K, $\gamma\sim-0.6$, but this
correction is only 0.2 at 30~K, and $\leq 0.1$ for $T_{\rm
dust}\geq60$~K. Hence, this temperature effect cannot explain the
observed variations of $\alpha$ across the L\,1157 area, and one has
to invoke variations of $\beta$.}. 
The frequency variation of dust opacity is not well known, and depends
on dust characteristics (see e.g.\ Visser et al.\ 1998, and references
therein). Interestingly, a larger $\beta$ -- as suggested in the flow
-- points toward smaller dust grains (e.g.\ Kr\"ugel \& Siebenmorgen
1994) and/or different chemical properties (e.g.\ Pollack et al.\
1994). Since the mm continuum emission in L\,1157 arises from shocked
regions, as shown by the morphological coincidences with shock tracers
such as NH$_3$ and SiO, this result would be consistent with models
predicting mantle evaporation (hence the rich chemistry observed in
the shocks of L\,1157) and even grain destruction (which may be the
only way to release silicon and produce strong SiO emission, Schilke
et al.\ 1997).

\begin{figure}[t]
\includegraphics[angle=270,width=8.8cm]{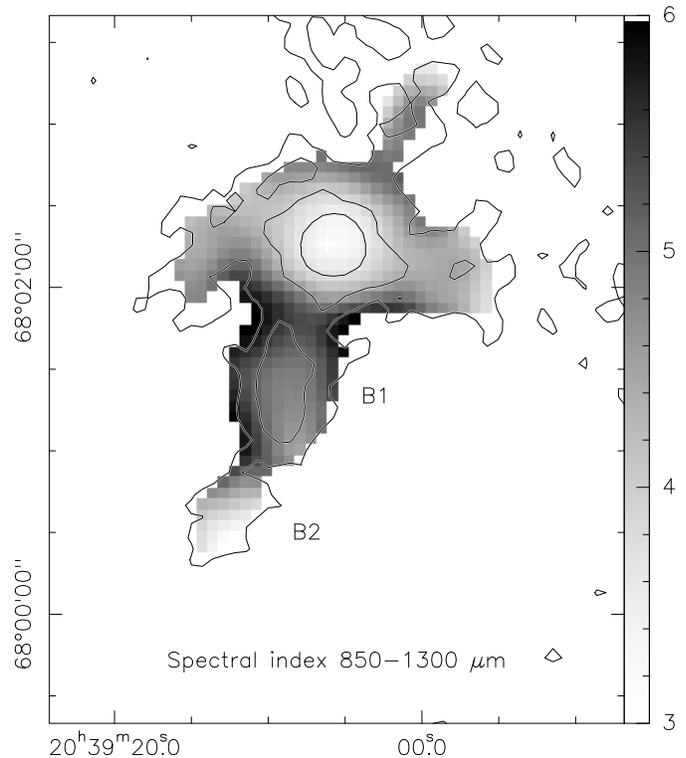} 
\caption{Grey scale map of the spectral index around the blueshifted
lobe of L\,1157, computed between 850~$\mu$m and 1.3~mm. The value of
the spectral index is indicated by the scale shown on the right. Three
contours of the 1.3\,mm emission map (Fig.\,1) are also drawn for
orientation. Before computing the index, we checked that both images
agreed in position within $\sim$2$''$, and smoothed them to the same
resolution ($20''$).}
\label{fig.spectralindex}
\end{figure}

\section{Conclusion} 

The observations presented in this letter demonstrate that the mm
continuum emission in L\,1157 originates from {\em four} distinct
components: the circumstellar disk, the infalling envelope, the dense
remnant of the cloud in which the protostar has formed, and the
outflow. The emission from the outflow represents $\sim$20\% of the
total flux at 1.3~mm, and is dominated by dust thermal emission.

The spectral index computed between 1.3~mm and 850~$\mu$m shows a
significant difference between the protostellar envelope and the
outflow emission. This could be related to an important line
contamination of the 850~$\mu$m map, and/or to variations of the dust
opacity law index. In the latter case, this may be an indication of
different dust properties, possibly of smaller grains in the
outflow. Observations of additional bipolar flows are needed to check
if such variations in the spectral index are a common phenomenon, and
to investigate their origin.

Finally, we caution that large-scale mm and submm continuum mapping of
molecular clouds (e.g.\ Motte et al.\ 1998) or of the vicinity of
protostellar objects (e.g.\ Shirley et al.\ 2000) can potentially be
confused by the presence of outflow emission, such as that from
L\,1157. Some emission knots could easily be misinterpreted as, e.g.,
starless clumps. The observations reported here demonstrate that
emission from bipolar outflows can be important, and must be taken
into account when interpreting the structure of mm maps of star
forming regions.

\acknowledgements {The authors are grateful to the IRAM staff at Pico
Veleta, to R.~Neri for valuable help with the observations, and to
Y.L.~Shirley and N.J.~Evans for providing a digital version of the
850~$\mu$m map. This research have been partially supported by Spanish
DGES grant AYA2000-927.}

\end{document}